\documentclass[twocolumn,prl,showpacs,superscriptaddress,preprintnumbers,amsmath,amssymb]{revtex4}
\usepackage[dvips]{graphicx}
\usepackage{dcolumn}
\usepackage{bm}
\begin{document}

\title{Determining the in-plane orientation of the ground-state orbital of CeCu$_2$Si$_2$}

\author{T. Willers}
  \affiliation{Institute of Physics II, University of Cologne,
   Z{\"u}lpicher Stra{\ss}e 77, D-50937 Cologne, Germany}
\author{F. Strigari}
  \affiliation{Institute of Physics II, University of Cologne,
   Z{\"u}lpicher Stra{\ss}e 77, D-50937 Cologne, Germany}
\author{N. Hiraoka}
  \affiliation{National Synchrotron Radiation Research Center (NSRRC), 101 Hsin-Ann Road, Hsinchu 30077, Taiwan}
\author{Y. Q. Cai}
  \affiliation{Photon Sciences, Brookhaven National Laboratory, Upton, New York 11973, USA}
\author{M. W. Haverkort}
	\affiliation{Max Planck Institute for Solid State Research, Heisenbergstra{\ss}e 1, D-70569 Stuttgart, Germany}
\author{K.-D. Tsuei}
  \affiliation{National Synchrotron Radiation Research Center (NSRRC), 101 Hsin-Ann Road, Hsinchu 30077, Taiwan}
\author{Y.F. Liao}
  \affiliation{National Synchrotron Radiation Research Center (NSRRC), 101 Hsin-Ann Road, Hsinchu 30077, Taiwan}
\author{S. Seiro}
  \affiliation{Max Planck Institute for Chemical Physics of Solids, N{\"o}thnizer Stra{\ss}e 40, D-01187 Dresden, Germany}
\author{C. Geibel}
  \affiliation{Max Planck Institute for Chemical Physics of Solids, N{\"o}thnizer Stra{\ss}e 40, D-01187 Dresden, Germany}
\author{F. Steglich}
  \affiliation{Max Planck Institute for Chemical Physics of Solids, N{\"o}thnizer Stra{\ss}e 40, D-01187 Dresden, Germany}
\author{L. H. Tjeng}
  \affiliation{Max Planck Institute for Chemical Physics of Solids, N{\"o}thnizer Stra{\ss}e 40, D-01187 Dresden, Germany}
\author{A. Severing}
  \affiliation{Institute of Physics II, University of Cologne,
   Z{\"u}lpicher Stra{\ss}e 77, D-50937 Cologne, Germany}
\date{\today}

\begin{abstract}
We have successfully determined the hitherto unknown sign of the $B_4^4$ Stevens crystal-field parameter of the tetragonal heavy-fermion compound CeCu$_2$Si$_2$ using vector $\hat{\textbf{q}}$ dependent non-resonant inelastic x-ray scattering (NIXS) experiments at the cerium N$_{4,5}$ edge. The observed difference between the two different directions $\hat{\textbf{q}}\|[100]$ and $\hat{\textbf{q}}\|[110]$ is due to the anisotropy of the crystal-field ground state in the (001) plane and is observable only because of the utilization of  higher than dipole transitions possible in NIXS. This approach allows us to go beyond the specific limitations of dc magnetic susceptibility, inelastic neutron scattering, and soft x-ray spectroscopy, and provides us with a reliable information about the orbital state of the $4f$ electrons relevant for the quantitative modeling of the quasi-particles and their interactions in heavy-fermion systems.

\end{abstract}


\maketitle

\begin{figure}
    \centering
    \includegraphics[width=1.0\columnwidth]{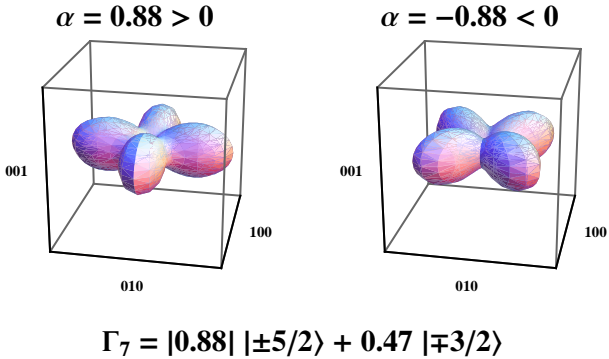}
    \caption{(color online) Angular distribution of two $\Gamma_7$ crystal field states that have the same $|\pm 5/2\rangle$ admixture ($\alpha^2=0.81$) but different signs of $\alpha$.}
    \label{Fig1}
\end{figure}

$4f$ heavy-fermion systems exhibit many fascinating phenomena including unconventional superconductivity and non-Fermi liquid behavior emerging in the vicinity of quantum critical points. In these systems the conduction band electrons are dressed by the hybridization with $4f$ electrons to form Landau quasi-particles with large effective masses. It is of great interest to identify the relevant quantum state of these $4f$ electrons in order to understand the formation and character of these quasi-particles and their interactions. Specifically, it is well known that the crystal-field split Hund's rule ground state of the unhybridized $4f$ electrons is highly anisotropic and it was recognized already early on that this anisotropy is important for the modeling of the properties of heavy-fermion systems~\cite{GunnarssonPRB42,ZwicknaglAdvPhys41}. There are examples for strongly anisotropic heavy fermi liquid formation~\cite{MatsumotoPRB84} and several authors have pointed out how the crystal-field anisotropy affects the band structure and shape of the Fermi surface~\cite{ZwicknaglPulst1993} and/or how it may influence the formation of the superconducting state~\cite{ThalmeierPRL101, Nevidomskyy_2009, FlintPRL105}.

However, for the largest group of heavy-fermion materials where the Ce$^{3+}$ (or Yb$^{3+}$) ion has tetragonal point symmetry (e.g. all CeM$_2$Si$_2$, CeM$_2$Ge$_2$, CeMIn$_5$, YbM$_2$Si$_2$, YbM$_2$Ge$_2$ with M being a transition metal, or CePt$_3$Si) the crystal-field ground-state is only incompletely determined because the orientation of the ground-state orbital in a symmetry higher than twofold cannot be determined with standard dipole techniques like inelastic neutron scattering, or more recently applied for crystal-field purposes, soft x-ray absorption spectroscopy. In this work we therefore utilize the higher than dipole transitions in non-resonant inelastic x-ray scattering (NIXS) \cite{LarsonPRL99,HaverkortPRL99,GordonEPL81,GordonProceedings2009,BradleyPRB81,CaciuffoPRB81,SenGuptaPRB84,BradleyPRB84,GordonJElecSpec184,HiraokaEPL96,vanderLaanPRL108} and introduce it as a spectroscopic tool new to this research field of heavy fermion physics. We will show below that we can resolve reliably the so-far unknown orientation of the ground-state orbital in the archetype system CeCu$_2$Si$_2$ which was the first heavy-fermion compound where unconventional superconductivity was discovered more than 30 years ago~\cite{SteglichPRL43}.

In tetragonal $4f$ systems the sixfold degenerate Hund's rule ground state of Ce$^{3+}$(J=5/2) is split into three Kramer's doublets under the influence of the crystal field. The crystal-field Hamiltonian can be written in Stevens approximation as $H_{CF}=B_2^0 O_2^0 + B_4^0 O_4^0 + B_4^4 O_4^4$ and the eigenfunctions can be represented in the basis of $|J_z \rangle$ when the fourfold symmetric tetragonal [001] axis is chosen as quantization axis. There are two $\Gamma_7$ doublets $\Gamma^1_7$ = $\alpha|\pm5/2\rangle + \sqrt{1-\alpha^2}|\mp 3/2 \rangle$ and $\Gamma^2_7$ = $\sqrt{1-\alpha^2} |\pm5/2\rangle - \alpha|\mp 3/2 \rangle$, and one $\Gamma_6$ which is a pure $|\pm1/2 \rangle$ doublet. The $\Gamma_6$ as a pure $|\pm1/2 \rangle$ has full rotational symmetry around [001] but the mixed $\Gamma_7$ states do not. Both $\Gamma_7$ states have a fourfold symmetry around [001] and for a given spatial distribution of the $4f$ wave function there are two solutions  which differ in their orientations within the (001) plane by 45$^\circ$ (see Fig.~1).  Which orientation applies to the ground state depends on the sign of $\alpha$. For $\alpha>0$ the wings of a $\Gamma_7$ ground state point along [100] and for $\alpha<0$ along [110]. The three Stevens parameters $B_2^0$, $B_4^0$, and $B_4^4$ are fully determined by the mixing parameter $\alpha$ and the energy level splittings $\Delta$$E_1$ and $\Delta$$E_2$, whereby the sign of $\alpha$ is related to the sign of $B_4^4$ such that $\alpha<0$ corresponds to $B_4^4>0$ and vice versa.

For CeCu$_2$Si$_2$ the crystal-field transition energies have been determined with inelastic neutron scattering \cite{GoremychkinPRB47} on polycrystalline samples. Inelastic neutron scattering is the ideal tool to determine the level splittings but the combination of phonon scattering and broad magnetic excitations in the same energy window prevents the unambiguous determination of the magnetic intensities. Therefore the ground-state wave function was determined with single crystal susceptibility measurements \cite{BatloggJApplPhys55,GoremychkinPRB47} and the crystal-field parameters which reproduce the susceptibility are given by Goremychkin \textit{et al.}~\cite{GoremychkinPRB47} and correspond to $|\alpha|=0.88$. Note, that the sign of $\alpha$ was not determined. Single crystal susceptibility measurements are a common way of determining wave functions in heavy fermion compounds. More recently, soft x-ray absorption spectroscopy with linear polarized light was shown to be a useful local probe to determine the anisotropy of the wave functions spectroscopically \cite{HansmannPRL100,WillersPRB80,WillersPRB81,WillersPRB85,WillersPRL107}. But both soft x-ray absorption spectroscopy and neutron scattering are governed by dipole transitions so that these methods are insensitive to anisotropies with a higher than twofold rotational axis and therefore cannot distinguish the two possible orientations of the $\Gamma_7$ orbital in the tetragonal (001) plane, which is synonym with not being able to determine the sign of $\alpha$ or $B_4^4$. In principle this in-plane anisotropy can be observed in direction dependent isothermal magnetization measurements, but for most cerium materials it would require even at 1~K magnetic fields of more than 20~T. Therefore, as for many other tetragonal heavy-fermions compounds, the $\Gamma_7$  ground state in the archetype heavy-fermion compound CeCu$_2$Si$_2$ could only be characterized by the absolute value of the mixing parameter $|\alpha|$ (or $|B_4^4|$) leaving the question about the orientation of the ground state within the (001) plane unanswered.

Here we present NIXS data of CeCu$_2$Si$_2$ single crystals and show how the higher multipole transitions can reveal the anisotropies of the ground-state wave-function which have not been accessible with other spectroscopies. NIXS has became feasible thanks to the high brilliance of modern synchrotrons and advanced instrumentation and has developed to a rapidly growing field providing complementary means to x-ray absorption methods in the study of core level excitations, but offering the possibility to go beyond the dipole limits \cite{LarsonPRL99,HaverkortPRL99,GordonEPL81,GordonProceedings2009,BradleyPRB81,CaciuffoPRB81,SenGuptaPRB84,BradleyPRB84,GordonJElecSpec184,HiraokaEPL96,vanderLaanPRL108}.
For rare earth materials higher multipole $4d\rightarrow 4f$ core level excitations (N$_{4,5}$-edge) have been observed for large momentum transfers $|\textbf{q}|$ by Gordon \textit{et al.}~\cite{GordonEPL81,GordonJElecSpec184} and Bradley \textit{et al.}~\cite{BradleyPRB84}. It was actually demonstrated experimentally how the dipole signal decreases with increasing $|$\textbf{q}$|$ while dipole forbidden transitions appear. These findings are in good agreement with simulations based on the local many body approach by Haverkort \textit{et al.}~\cite{HaverkortPRL99} and it was suggested that vector $\hat{\textbf{q}}$ dependent NIXS experiments with large momentum transfers on single crystals should have the potential to give insight into the ground state symmetry.
The direction of the $\hat{\textbf{q}}$ vector with respect to a crystallographic axis is the analog of the polarization in an x-ray absorption experiment with linear polarized light. Gordon \textit{et al.} performed first NIXS experiments on cubic single crystals of MnO and CeO$_2$ at the Mn M$_{2,3}$ and Ce N$_{4,5}$ edges and found some differences for different $\hat{\textbf{q}}$ directions in the higher multipole scattering. This has been interpreted as sensitivity to anisotropies in charge densities with higher than two-fold symmetry~\cite{GordonProceedings2009}.

\begin{figure}
    \centering
    \includegraphics[width=1.0\columnwidth]{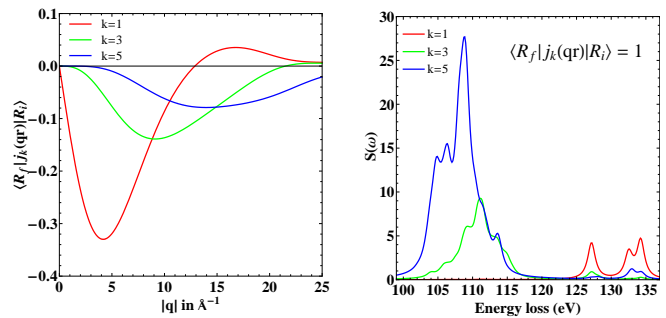}
    \caption{(color online) Left: k$^{th}$ order term of the radial part of $S(q,\omega)$  as function of momentum transfer. Right: k$^{th}$ order contribution of angular part of the scattering function expressed in terms of $S(\omega)$ versus energy transfer for $\left\langle R_f|j_k(\textbf{qr})|R_i \right\rangle$ = 1 (see Eq.~\ref{Scatt}).}
    \label{Fig2}
\end{figure}

\begin{figure}
    \centering
    \includegraphics[width=1.0\columnwidth]{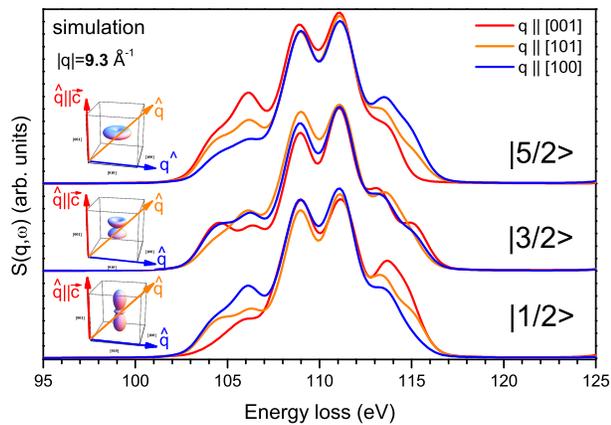}
    \caption{(color online) Simulations: Comparison of \textsl{in-plane} and \textsl{out-of-plane} scattering function $S(\textbf{q},\omega)$  of pure $J_z$ states, in-plane $\hat{\textbf{q}}\|[100]$ (blue),
     out-of-plane $\hat{\textbf{q}}\|[001]$ (red), and in between $\hat{\textbf{q}}\|[101]$ (orange).
     The calculations are done for $|\textbf{q}|=$9.3~\AA$^{-1}$ and are convoluted with a Lorenzian with FWHM = 0.3~eV and a Gaussian with FWHM = 1.32~eV.}
    \label{Fig3}
\end{figure}

\begin{figure}
    \centering
    \includegraphics[width=1.0\columnwidth]{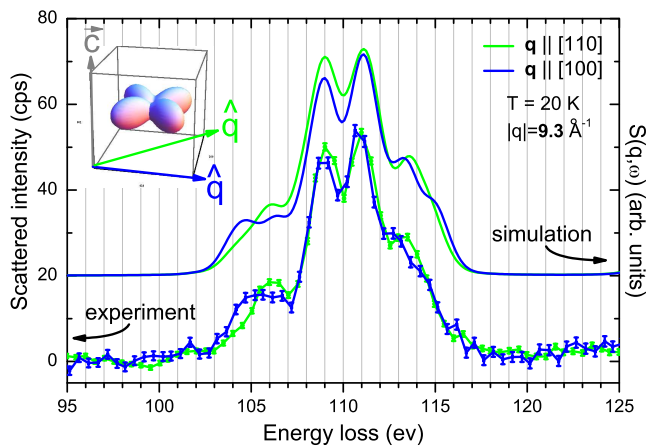}
    \caption{(color online) Simulation (top): Comparison of the scattering function $S(\textbf{q},\omega)$ for
     two \textsl{in-plane} directions $\hat{\textbf{q}}\|[100]$ (blue)			 and $\hat{\textbf{q}}\|[110]$ (green) assuming a $\Gamma_7$ ground state  with $\alpha<0$. The calculations are convoluted with a Lorenzian with FWHM = 0.3~eV
      and a Gaussian with FWHM = 1.32~eV.  Experiment (bottom): NIXS data of two CeCu$_2$Si$_2$ single
       crystals. The blue dots and lines corresponds to the set-up for $\hat{\textbf{q}}\|[100]$ and
       the green ones to $\hat{\textbf{q}}\|[110]$. The error bars reflect the statistical error.
       The lines are guides to the eye.}
    \label{Fig4}
\end{figure}

For the sake of clarity, we first review briefly the theoretical description of inelastic x-ray scattering~\cite{SchulkeBook,SoininenPRB72,HaverkortPRL99,GordonEPL81,CaciuffoPRB81,BradleyPRB84,BradleyPRL105,vanderLaanPRL108} and show later how the crystal-field affects the scattering cross section. The double differential cross section is the product of the Thomson photon cross section $\left(\frac{d\sigma}{d\Omega}\right)_{Tho}$ and the dynamical structure factor $S(q,\omega)$, the latter containing the physics of the material under investigation:
\begin{equation}
\frac{d^2\sigma}{d\Omega d\omega} = \left(\frac{d\sigma}{d\Omega}\right)_{Tho}  S(q,\omega)
\end{equation}
The dynamical structure factor is a function of the  scattering vector $\bf{q}=\bf{k_i}-\bf{k_f}$  and the energy loss $\omega=\omega_i-\omega_f$
\begin{eqnarray}
S(\textbf{q},\omega) = \sum_{f}|\left\langle f|e^{i\textbf{qr}}|i \right\rangle |^2 \delta(\hbar\omega_i-\hbar\omega_f-\hbar\omega).\nonumber
\end{eqnarray}
Here $i$ and $f$ are the initial and final states. In order to distinguish between dipole, octupole etc. terms the transition operator $e^{i\textbf{qr}}$ is expanded in semi-normalized (Racah's normalization) spherical harmonics $C^{\hat{\textbf{q}}^*}_{km}$ and $C^{\hat{\textbf{r}}}_{km}$.
This results in a sum over spherical Bessel functions $j_k(\textbf{qr})$ and the wave functions can be factorized  into a radial and angular part so that S(\textbf{q},$\omega$) can be written as
\begin{eqnarray}
&S(\textbf{q},\omega) =  \qquad\qquad\qquad\ \qquad\qquad\qquad\ \qquad\qquad\qquad\ & \nonumber\\
&\sum\limits_{f}\left| \sum\limits^{\infty}_{k}i^k(2k+1) \left\langle R_f|j_k(\textbf{qr})|R_i \right\rangle \sum\limits^{k}_{m=-k}\left\langle \phi_f|C^{\hat{\textbf{q}}^*}_{km} C^{\hat{\textbf{r}}}_{km}|\phi_i \right\rangle\right|^2 &  \nonumber\\
&\qquad\qquad\qquad \times \delta(\hbar\omega_i-\hbar\omega_f-\hbar\omega) &
\label{Scatt}
\end{eqnarray}
Due to  the triangle condition and parity selection rules only terms with k=1(dipole), 3(octupole), and 5(triakontadipole) contribute to the Ce $4d\rightarrow 4f$ (N$_{4,5}$) transitions. The radial part $\left\langle R_f|j_k(\textbf{qr})|R_i \right\rangle$ of the wave functions have been calculated within the Hartree-Fock approximation using Cowan's code~\cite{CowanBook} and the k$^{th}$ contributions are shown as function of momentum transfer $|\textbf{q}|$ in the left panel of  Fig.~\ref{Fig2}. For moderate magnitudes of $|\bf{q}|$ the radial part is dominated by dipole scattering, but already at 5~\AA$^{-1}$ octopole scattering is non-negligible, and at even higher momentum transfers the scattering is dominated by the higher multipoles. This behavior is commonly called q-dependent multipole selection rules. The right panel of Fig.~\ref{Fig2} shows the k$^{th}$ order of the angular part as function of energy loss.
Higher multipoles have different selection rules so that extra intensity at different energy losses becomes visible in the angular part when at large $|\textbf{q}|$ octupole and triakontadipole transitions take place.


Having described the $|\bf{q}|$ dependence of the NIXS intensities, we now turn to the vector $\hat{\textbf{q}}$ dependence which is at the heart of our study. We first of all evaluate the sensitivity of NIXS at the N$_{4,5}$ edge to crystal-field anisotropies in general by comparing simulated spectra for different directions of $\hat{\textbf{q}}$. These crystal-field anisotropies are included when the initial and final states in Eq.~\ref{Scatt} are eigenstates of a Hamiltonian that contains in addition to the atomic Coulomb and spin-orbit interactions also non-vanishing crystal-field terms.
We note that the interference terms which vanish if the angular intensity is integrated over all directions $\hat{\textbf{q}}$ \cite{BradleyPRB84,vanderLaanPRL108} are included in our calculations. Figure~\ref{Fig3} shows the simulation of $S(\textbf{q},\omega)$ for large momentum transfers for the three pure $J_z$ states, considering some realistic lifetime broadening and width due to instrumental resolution (see below). The pure states have rotational symmetry so that we compare the in-plane scattering ($\hat{\textbf{q}}\|[100]$) with scattering out-of-plane, i.e. for $\hat{\textbf{q}}\|[001]$, and some direction in between ($\hat{\textbf{q}}\|[101]$). There is a clear direction dependence, so that the different $J_z$ states are distinguishable. This is similar to the in-plane/out-of-plane polarization dependence in soft x-ray absorption at the cerium M$_{4,5}$ edge \cite{HansmannPRL100,WillersPRB80}.

The next step is to investigate the in-plane anisotropy of a mixed $\Gamma_7$ state as shown in  Fig.~\ref{Fig1}. For that we calculate and compare $S(\textbf{q},\omega)$ with $\alpha <0 $ for the two in-plane directions  $\hat{\textbf{q}}\|[100]$ and $\hat{\textbf{q}}\|[110]$ and we end up with distinguishable spectra (see top of Fig.~\ref{Fig4}). This clearly shows that such an experiment can be used to determine the sign of $\alpha$. The simulation in Fig.~\ref{Fig4} assumes a negative value for $\alpha$ and the corresponding orbital orientation is shown in the inset. For a 45$^\circ$ rotation around the $c$-axis [001], i.e. for a positive value of $\alpha$ the spectra are inverted. We found that these in-plane spectra are fairly insensitive to the precise value of $\alpha$ as long as $\alpha$ is neither zero nor one. In the latter case the orbital would have full rotational symmetry around [001] and consequently  $S(\textbf{q},\omega)$ would look the same for both in-plane directions.

To determine the orbital orientation in CeCu$_2$Si$_2$ we performed Ce N$_{4,5}$ NIXS measurements at the IXS end station of the Taiwan beamline BL12XU at SPring8 on two CeCu$_2$Si$_2$ single crystals. One crystal had a polished (001) and the other one a polihed (110) surface so that $\hat{\textbf{q}}\|[100]$ and $\hat{\textbf{q}}\|[110]$ could be realized in a specular geometry. The crystals were cooled down to about 20~K in a closed cycle cooler so that only the ground state was populated. The incident photon energy of 9890~eV and a scattering angle of 2$\Theta$=138$^o$ correspond to a momentum transfer of $|\textbf{q}| = 9.3$ \AA$^{-1}$. The energy resolution as determined from the full width at half maximum (FWHM) of the elastic line was 1.32~eV.  The signals of nine spherically bent backscattering Si(555) analyzers at two meter distance from the sample were combined in order to gain higher count rates. The energy loss was determined from the energy position of the elastic line measured before and after each set of spectra.

The measured Ce N$_{4,5}$ spectra are shown at the bottom of Fig.~\ref{Fig4}. Only a linear background has been subtracted. Comparing our single crystal data with the polycrystalline results by Gordon \textit{et al.} confirms the trivalent character of the cerium ion. We also find, similarly  to Gordon \textit{et al.}, that the experimental dipole transitions are much broader than the higher multipole transitions. In Ref.~\cite{SenGuptaPRB84,GordonJElecSpec184} it was reasoned that this is due to interactions of the dipole final states with the continuum.

The two measured directions in our single crystal experiment can be clearly distinguished. The experimental spectra agree very well with our theoretical predictions. In particular, the vector $\hat{\textbf{q}}$ dependence is in good agreement with the experiment. For example the inversion of the anisotropy at around 105 and 113~eV, as predicted in the calculations, is well reproduced.
Here we note that for the calculation the Hartree-Fock values of the Slater integrals were reduced to about 68\% for the $4f$-$4f$ Coulomb interactions and to about 88\% for the $4d$-$4f$ interactions to reproduce best the energy positions of the main spectral features. These values compare well to a 60\% to 80\% scaling of the atomic values  needed to describe isotropic rare earth M$_{4,5}$ ($3d\rightarrow 4f$) x-ray absorption spectra and account for configuration interaction effects not included in the Hartree-Fock scheme~\cite{TholePRB32,TanakaJPSC63}. The simulated spectra were convoluted with a Lorentzian and a Gaussian with a FWHM=0.3~eV and 1.32~eV, respectively to account for the intrinsic line width and experimental resolution. 
This experiment shows clearly that in CeCu$_2$Si$_2$ $\alpha$ is negative so that the wings of the Ce$^{3+}$  ground-state orbital point in [110] direction.

In conclusion, our simulation show that vector $\hat{\textbf{q}}$ dependent NIXS at the cerium N$_{4,5}$-edge in the limit of high $|\textbf{q}|$ is sensitive to the crystal-field ground-state symmetry. In addition, we show that the  higher multipole character of the scattering at large momentum transfers gives insight into anisotropies which are higher than twofold so that the two possible orientations of a $\Gamma_7$ orbital in tetragonal symmetry can be distinguished. We have verified this experimentally by performing a NIXS experiment on CeCu$_2$Si$_2$ single crystals. The spectral line shape and the observed differences between the two measured directions $\textbf{q}\|[100]$ and $\textbf{q}\|[110]$ are well explained by atomic full multiplet calculations. These unambiguously show that the ground state orbital in CeCu$_2$Si$_2$ is of the type $\Gamma_7 = -|\alpha| |\pm5/2\rangle + \sqrt{1-\alpha^2}|\mp 3/2 \rangle$. Thus, $\textbf{q}$ dependent NIXS provides information which is not accessible with dc susceptibility, neutron scattering  and/or soft x-ray absorption spectroscopy and  thereby completes the spectroscopic toolbox  for determining the crystal-field ground-states in tetragonal cerium compounds.
It is now possible and will be of great interest to investigate systematically the orientation of the ground-state orbitals of other tetragonal, heavy-fermion cerium compounds. 
Also, NIXS with hard x-rays has the advantage of a large penetration depth so that ultra high vacuum is obsolete, which allows the use of advanced sample environments for pressure and/or low temperatures. 

This work was supported by the DFG through project AOBJ 583872. The experiment was performed in BL12XU/SPring-8 under approvals with JASRI (No. 2011A4254) and NSRRC, Taiwan (2011-2-053). YQC is supported by the U.S. Department of Energy, Office Basic Energy Science, under the Contract No. DE-AC02-98CH10886.



\end{document}